\documentclass[%
 reprint,
superscriptaddress,
preprintnumbers,
nofootinbib,
 amsmath,amssymb,
 aps,
]{revtex4-2}

\usepackage{graphicx}
\usepackage{dcolumn}
\usepackage{bm}
\usepackage{hyperref}

\hypersetup{
    unicode=false,          
    pdftoolbar=true,        
    pdfmenubar=true,        
    pdffitwindow=false,     
    pdfstartview={FitH},    
    pdftitle={},    
    pdfauthor={},     
    pdfsubject={},   
    pdfcreator={},   
    pdfproducer={}, 
    pdfkeywords={} {} {}, 
    pdfnewwindow=true,      
    colorlinks=true,       
    linkcolor=blue, 
    citecolor=blue,        
    filecolor=magenta,      
    urlcolor=blue,           
	breaklinks=true
} 

\usepackage{amssymb}
\usepackage{amsmath}
\usepackage{amsfonts}
\usepackage{xspace}
\usepackage{bm,bbm}
\usepackage{relsize}
\usepackage{siunitx}
\usepackage{xcolor}
\usepackage{upgreek}

\makeatletter
\DeclareFontFamily{OMX}{MnSymbolE}{}
\DeclareSymbolFont{MnLargeSymbols}{OMX}{MnSymbolE}{m}{n}
\SetSymbolFont{MnLargeSymbols}{bold}{OMX}{MnSymbolE}{b}{n}
\DeclareFontShape{OMX}{MnSymbolE}{m}{n}{
    <-6>  MnSymbolE5
   <6-7>  MnSymbolE6
   <7-8>  MnSymbolE7
   <8-9>  MnSymbolE8
   <9-10> MnSymbolE9
  <10-12> MnSymbolE10
  <12->   MnSymbolE12
}{}
\DeclareFontShape{OMX}{MnSymbolE}{b}{n}{
    <-6>  MnSymbolE-Bold5
   <6-7>  MnSymbolE-Bold6
   <7-8>  MnSymbolE-Bold7
   <8-9>  MnSymbolE-Bold8
   <9-10> MnSymbolE-Bold9
  <10-12> MnSymbolE-Bold10
  <12->   MnSymbolE-Bold12
}{}

\let\llangle\@undefined
\let\rrangle\@undefined
\DeclareMathDelimiter{\llangle}{\mathopen}%
                     {MnLargeSymbols}{'164}{MnLargeSymbols}{'164}
\DeclareMathDelimiter{\rrangle}{\mathclose}%
                     {MnLargeSymbols}{'171}{MnLargeSymbols}{'171}
\makeatother

\renewcommand{\vec}[1]{\bm{#1}}

\newcommand{\norm}[1]{\lVert#1\rVert}

\begin{document}

\title{Classifying topology in photonic crystal slabs with radiative environments}

\author{Stephan Wong}
\email[Email: ]{stewong@sandia.gov}
\affiliation{Center for Integrated Nanotechnologies, Sandia National Laboratories, Albuquerque, New Mexico 87185, USA}

\author{Terry A. Loring}
\affiliation{Department of Mathematics and Statistics, University of New Mexico, Albuquerque, New Mexico 87131, USA}

\author{Alexander Cerjan}
\affiliation{Center for Integrated Nanotechnologies, Sandia National Laboratories, Albuquerque, New Mexico 87185, USA}

\date{\today}

\begin{abstract}
In the recent years, photonic Chern materials have attracted substantial interest as they feature topological edge states that are robust against disorder, promising to realize defect-agnostic integrated photonic crystal slab devices.
However, the out-of-plane radiative losses in those photonic Chern slabs has been previously neglected, yielding limited accuracy for predictions of these systems' topological protection.
Here, we develop a general framework for measuring the topological protection in photonic systems, such as in photonic crystal slabs, while accounting for in-plane and out-of-plane radiative losses.
Our approach relies on the spectral localizer that combines the position and Hamiltonian matrices of the system to draw a real-picture of the system's topology.
This operator-based approach to topology allows us to use an effective Hamiltonian directly derived from the full-wave Maxwell equations after discretization via finite-elements method (FEM), resulting in the full account of all the system's physical processes.   
As the spectral FEM-localizer is constructed solely from FEM discretization of the system's master equation, the proposed framework is applicable to any physical system and is compatible with commonly used FEM software.
Moving forward, we anticipate the generality of the method to aid in the topological classification of a broad range of complex physical systems.
\end{abstract}

\maketitle


\section{Introduction}

Originally discovered in the context of electronic systems, the concept of topological insulators has been generalized to photonic structures thanks to the platform-independent framework of topological band theory used to classify such systems.
Subsequently, over the past decade, there has been substantial interest in photonic topological insulators due to their potential to yield next-generation optical devices based on their topologically protected edge states~\cite{Ozawa2019}. 
For example, non-reciprocal waveguiding modes can be achieved in photonic crystals exhibiting non-trivial topology from broken time-reversal symmetry, which can be realized using gyro-electric or gyro-magnetic materials~\cite{Wang2009, Bahari2017} as well as in driven nonlinear systems~\cite{he_floquet_2019}. Similar waveguiding modes can also be found in a variety of metamaterials, such as those based on shifted ring-resonator arrays~\cite{Hafezi2011, Hafezi2013}, helical waveguide arrays~\cite{Rechtsman2013}, or that use synthetic dimensions~\cite{yuan_achieving_2015, Lustig2019, dutt_experimental_2019, dutt_single_2020}.
Moreover, using solely the crystalline symmetries of the photonic crystal, different classes of non-trivial topology can also be attained, leading to robust waveguiding states along bends that preserve the crystalline symmetry~\cite{Wu2015, Parappurath2020, Xue2021, Ma2016, Wong2020} or robust cavity-like states for enhanced light-matter interactions~\cite{Ota2020, Kim2020, Kruk2021}.

\begin{figure}[!]
\center
\includegraphics[width=\columnwidth]{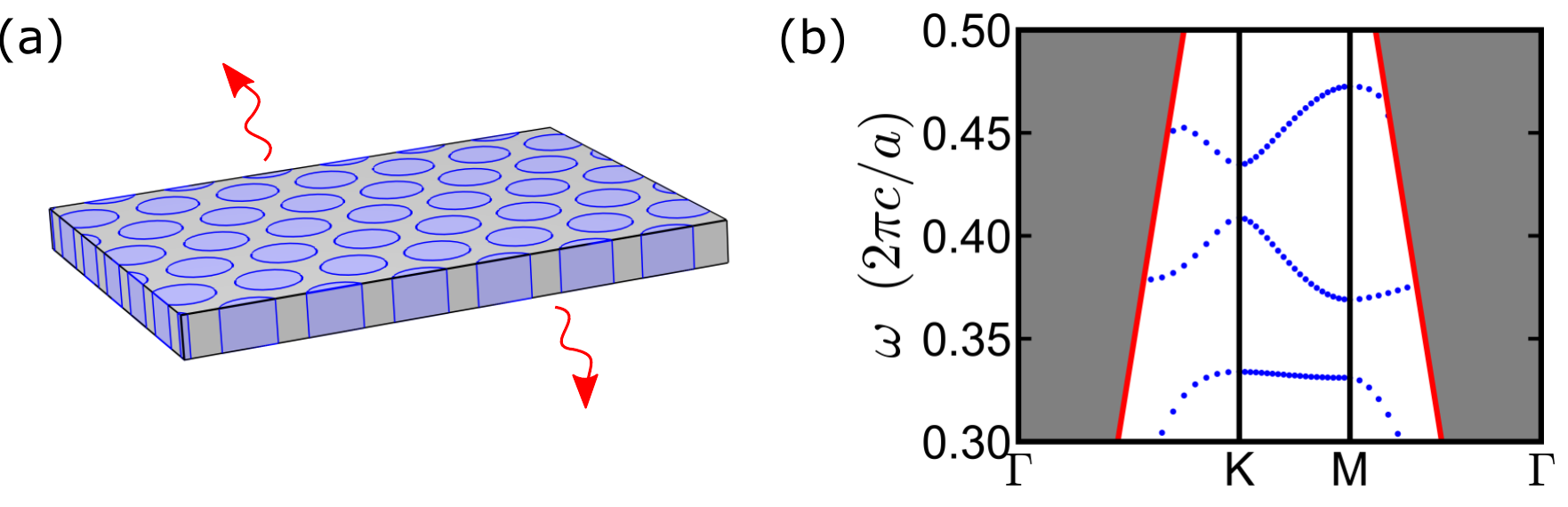}
\caption{
\textbf{Gapless environment for photonic crystal slabs.}
(a) Schematic of a free-standing photonic crystal slab in a three-dimensional (3D) geometry. 
Photonic structures are inherently 3D and are usually surrounded by an homogeneous material that features a light cone. As such, out-of-plane radiative losses, depicted by the red arrows, are inherent to such structures.
The photonic crystal is a triangular lattice with lattice constant $a=\SI{1}{\micro\metre}$ composed of dielectric rods, $\bar{\epsilon}_{jj} = 14$ for $j=x,y,z$, of radius $r=0.37a$ and  height $t = 0.5a$ embedded in a gyro-electric material slab, $\bar{\epsilon}_{jj} = 1$ and $\bar{\epsilon}_{xy} = -0.4i$, of thickness $t = 0.5a$.
(b) Band structure of the photonic crystal slab in (a) over the first Brillouin zone, for a transverse magnetic-like polarization. 
The photonic band gap at around $\omega = 0.42 [ 2\pi c/a]$ is the ``topological band gap'' known from extrapolation from the two-dimensional photonic crystal approximation. The shaded region depicts those frequencies and wavevectors that are at, or above, the light line of the surrounding air. The red line depicts the light line, $\omega = c|k|$. Above the light line, photonic crystal slabs generally exhibit resonances, not bound states.
}
\label{fig:slab}
\end{figure}

Material topology in electronic systems is identified through the system's band structure and Bloch eigenstates using invariants defined on the system's Brillouin zone; similarly, topological robustness is defined in terms of the system's bulk band gap~\cite{Hasan2010, Chiu2016}.
Traditionally, these same invariants have been used to classify topology in photonic systems as well. However, the analogy between topological insulators in electronic systems and in photonic systems is not always exact.
For realizations of photonic topological insulators operating at longer wavelengths~\cite{Wang2009, Poo2011, Cheng2016}, the system can be bounded by materials that provide excellent approximations of perfect electric conductors~\cite{Rylander2013}, yielding similar open boundary conditions to those that appear in electronic systems.
In contrast, photonic topological insulators operating at technologically relevant wavelengths and length-scales are generally based on photonic crystal slab motifs, and border free space on at least one surface, which is gapless above the light cone [Fig.~\ref{fig:slab}]. 
This gapless radiative environment has two relevant consequences for the classification of photonic topology: (1) For those wavevectors above the light-cone, the photonic crystal slab does not possess a true band structure consisting entirely of bound state solutions with real frequencies and instead only exhibits a resonance band structure characterized by complex frequencies. As such, it is not known whether standard topological invariants, defined by integrating over a system's occupied bands, can be meaningfully applied to the resonances of such photonic slab systems. (2) Similarly, the full system of photonic crystal slab plus surrounding environment is gapless above the light-cone, meaning that according to band theory the system's topological protection is not well-defined.
Thus, as already noted by Raghu and Haldane~\cite{Haldane2008}, the radiative environment has traditionally limited the notion of topological protection---as defined by topological band theory---in photonic topological insulators.
Therefore, although topological band theory can be applied to photonic systems to yield some insight into their topological properties, this cannot be viewed as a complete picture as the setting goes beyond the scope of band theoretic approaches; any complete picture of topological photonic crystal slabs must account for their inherent out-of-plane radiative losses, and provide a definition of topological protection despite these losses.

Here, we present a general framework for classifying topology in realistic three-dimensional (3D) photonic systems that directly accounts for both in-plane and out-of-plane radiative losses. To do so, we solve two interconnected challenges, demonstrating how to perform dimensional reduction to calculate invariants of 2D systems, such as Chern numbers, for photonic crystal slab systems that are inherently 3D, and showing how this generalizes to non-Hermitian systems so that radiative losses can be properly accounted for.
The framework we develop is rooted in the spectral localizer, which combines the position matrices and the Hamiltonian matrix of the system to draw a local picture of a system's topology~\cite{Loring2015}.
Using the operator-based approach of the spectral localizer, we construct a spectral FEM-localizer built on an effective Hamiltonian directly derived from the full-wave Maxwell equations via finite-element method (FEM).
We demonstrate the utility of the FEM-localizer framework through two fundamental examples in topological photonics: a 2D photonic Chern insulator and a 2D photonic Chern quasicrystal, proving that the spectral FEM-localizer approach is applicable to aperiodic structures that lie beyond the scope of topological band theory.
Finally, we apply the spectral FEM-localizer to study the topology of a photonic Chern slab, showing how to generate a local strong 2D invariant for the 3D system while directly accounting for the out-of-plane radiative losses.
Given the wide variety of systems that can be faithfully approximated by FEMs, we anticipate that our framework has broad applicability both within photonics and beyond, and will be useful for the study of the topology in complex physical systems where a band theoretic picture is not available.


\section{Overview of the spectral localizer}

Despite being a real-space approach to topology, the spectral localizer shares some conceptual similarities with topological band theory. A modern understanding of traditional band theoretic approaches to topology can be built from the concept of atomic limits---the limit in which the couplings between adjacent atoms, molecules, or structural decorations are turned off, and the system's band structure becomes completely flat. Atomic limits are topologically trivial as they always possess a complete symmetry-preserving localized Wannier basis~\cite{Klitzing1980, Kane2005}. Moreover, different systems are topologically equivalent if one system can be smoothly deformed (i.e., path continued) into the other system without closing the relevant bulk band gap (i.e., the band gap at the frequency of interest) or breaking any necessary symmetry. Thus, from this perspective, the question of material topology becomes one of whether a system can be path continued to an atomic limit; if so, it is trivial. Band theoretic approaches offer a few different methods for making this determination, either using standard topological invariants~\cite{Chiu2016}, or by comparing a system's band structure against the possible elementary band representations~\cite{Bradlyn2017}.

In contrast, the spectral localizer framework has emerged as a method to diagnose a system's topology from the real-space perspective of the atomic limit. The key idea is that just as the wavevector-space description of atomic limits is as a material with completely flat bands~\cite{Kitaev2009}, a real-space description of atomic limits can be understood in terms of the system's position matrices $X_j$ and Hamiltonian $H$ via the commutation relations 
\begin{equation}
\label{eq:ti_commutator}
[X_j^{(\textrm{AL})}, H^{(\textrm{AL})}] = 0, \quad j = 1, \ldots, d
,
\end{equation}
with $d$ the dimension of the system. In other words, in an atomic limit, $H^{(\textrm{AL})}$ is block diagonal, with each block corresponding to each decoupled atom, molecule, or structural element. Likewise, the position of each such interior degree of freedom is condensed into a single location. Hence, $X_j^{(\textrm{AL})}$ and $H^{(\textrm{AL})}$ commute for atomic limits.

From this real-space perspective, the question of topology then becomes one of understanding whether a system's non-commuting $H$ and $X_j$ matrices are nevertheless path continuable to commuting matrices, while preserving any necessary symmetry and maintaining the relevant bulk spectral gap. To perform this assessment directly in real-space for a $d$-dimensional Hermitian system in any of the ten Altland-Zirnbauer symmetry classes~\cite{Altland1997, Schnyder2008, Kitaev2009, Ryu2010}, one first forms the system's spectral localizer by combining $X_j$ and $H$ using a Clifford representation~\cite{Loring2015},
\begin{multline}
\label{eq:localizer}
L_{(x_1,\dots,x_d,E)}(X_1, \dots, X_d, H) = \\
\sum_{j=1}^d \kappa (X_j - x_j I) \otimes \Gamma_j + (H - EI) \otimes \Gamma_{d+1}
.
\end{multline}
Here, 
$\Gamma_1, \dots, \Gamma_{d+1}$ are $(d+1)$-dimensional Clifford representation satisfying $\Gamma_j^\dagger = \Gamma_j$, $\Gamma_j^2 = I$, and $\Gamma_j \Gamma_l = -\Gamma_l \Gamma_j$ for $j \ne l$, while $I$ is the identity matrix.
The spectral localizer is an inherently local approach to material topology; in Eq.~\eqref{eq:localizer}, $(x_1, \dots, x_d, E) \equiv \vec{\lambda}$ is the spatial coordinate $(x_1, \dots, x_d)$ and energy $E$ where the system's topology is being probed.
Finally, $\kappa$ is a hyperparameter chosen to make the units consistent between the position and Hamiltonian matrices, and to additionally balance the emphasis on the system's position information relative to its Hamiltonian. Typically, for gapped systems, it suffices to choose $\kappa \sim 2 g / L$~\cite{Loring2020, Dixon2023, Cerjan2023b} where $g$ is the width of the bulk band gap and $L$ is the length of the finite system considered.
Extensive studies have demonstrated the spectral localizer's versatile applicability across a broad range of topological systems, even beyond the traditional topological band theory~\cite{Loring2015, Fulga2016, Loring2017, Loring2020, Cerjan2022, Cerjan2022a, Liu2023, Franca2023, Franca2023a, Cerjan2023, Cheng2023, Dixon2023, Wong2023, Cerjan2023b, Chadha2024, Ochkan2024}. 

Using results from the study of $C^*$-algebras~\cite{Loring2015, Loring2017, Loring2020}, the spectrum of the spectral localizer has been proven to be connected to local topological markers for every discrete symmetry class (i.e., Altland-Zirnbauer class~\cite{Altland1997, Schnyder2008, Kitaev2009, Ryu2010}) for every physical dimension.
For a 2D Hermitian system in class A, namely a system lacking any discrete symmetries, the appropriate local topological invariant is the local Chern number, defined as 
\begin{equation}
C_{(x,y,E)}^{\text{L}}(X,Y,H) = \tfrac{1}{2} \textrm{sig} \big[ L_{(x,y,E)}(X,Y,H) \big],
\end{equation}
where $\text{sig}$ denotes the matrix's signature, the difference between its total number of positive and negative eigenvalues.
A non-zero local Chern number at $(x,y,E)$ indicates that the re-centered position $(X - xI)$, $(Y-yI)$ and Hamiltonian $(H-EI)$ matrices cannot be path continued to be commuting, meaning that the system cannot be deformed to the atomic limit.
In other words, a non-zero local Chern number, $C_{(x,y,E)}^{\text{L}} \neq 0$ tells us that the system is topologically distinct from the atomic limit, and is locally topologically non-trivial at the spatial and energy coordinate $(x,y,E)$.

Separate from its ability to identify material topology, the spectral localizer can be seen as a tool to identify localized states by looking at an approximate state $\vec{\psi}$ of both the position and Hamiltonian matrices of the system
\begin{equation}
\label{eq:joint_spectrum}
X_j \vec{\psi} \approx x_j \vec{\psi} \quad \text{and} \quad H \vec{\psi} \approx E \vec{\psi}
,
\end{equation}
In particular, the localizer gap, defined as the smallest singular value of the spectral localizer
\begin{multline}
\label{eq:localizer_gap}
\mu^\text{C}_{(x_1,\dots,x_d,E)}(X_1, \dots, X_d, H) = \\ 
\min \big[ | \sigma(L_{(x_1,\dots,x_d,E)}(X_1, \dots, X_d, H)) | \big]
,
\end{multline}
with $\sigma(L)$ being the set of eigenvalues of $L$,
gives a measure about large of a modification of the system is needed to realize such state $\vec{\psi}$.
As such, a localizer gap closing $\mu^\text{C}_{(x_1,\dots,x_d,E)}=0$ indicates that there exists such an approximate state $\vec{\psi}$ at energy $E$ that is localized at spatial position $(x_1, \ldots, x_d)$.

Altogether, the local topological markers and the localizer gap give a consistent picture of the topology locally in space and energy.
The local topological markers are used to probe the topology locally at space-energy coordinate $(x_1, \ldots, x_d, E)$, and cannot change as long as the localizer gap does not close ($\mu^\text{C}_{(x_1,\dots,x_d,E)} \neq 0$).
When the local markers change across some spatial or energy path from one topological phase to another topological phase, the localizer gap must close, resulting in a localized state [Eq.~\eqref{eq:joint_spectrum}] at the interface between the two topological distinct phases: this is precisely bulk-edge correspondence.

The spectral localizer is not the only real-space theory of material topology that provides local topological markers \cite{Kitaev2006, bianco_local_chern_2011, Fulga_et_al_scattering_top_ins, mitchell_amorphous_2018, Varjas_et_al_2020_KMP_method, prodan2011disordered, caio_topological_2019, mondragon-shem_robust_2019, hannukainen_local_2022, munoz-segovia_structural_2023, kim_replica_2023}. However, it is the only currently known theory of topology that preserves system sparsity, i.e., if $H$ is sparse, $L_{(x_1,\dots,x_d,E)}$ is sparse. Thus, unlike other local markers that typically require projecting into an occupied subspace, yielding still-relatively-large dense matrices, the calculation of local markers using the spectral localizer can leverage advances in sparse matrix algorithms to realize substantial numerical speedups. For example, finding a sparse matrix's signature does not require finding any of its eigenvalues, and can instead be found using Sylvester's law of inertia \cite{sylvester_xix_1852, higham2014sylvester}.


\section{Building the spectral FEM-localizer}

The spectral localizer is used to study a system's topology directly from its equations of motion, i.e., its master equation.
For example, it is straightforward to apply the spectral localizer framework to tight-binding models~\cite{Loring2015, Fulga2016, Cerjan2022, Liu2023, Franca2023, Franca2023a, Cerjan2023, Cheng2023, Wong2023, Chadha2024, Ochkan2024} where the position operators are diagonal matrices with entries being the spatial position $(x_1, \ldots, x_d)$ of the model's sites, and $H$ the tight-binding Hamiltonian.
However, the spectral localizer is not limited to such approximate descriptions of physical system.
Instead, so long as a system admits a discretization of some form (or some other method for generating a bounded matrix description of the system), the matrix of its discretized master equation can be inserted into the spectral localizer [Eq.~\eqref{eq:localizer}], where practical parameters and geometry can be used. 
With such a choice of discretization, the entries of the position matrices $X_j$ can then be chosen as the grid mesh positions from discretization of the master equation and the Hamiltonian matrix $H$ can be chosen as being any matrix whose eigenvector is a solution of the master equation, namely a matrix whose eigenproblem is physically meaningful in order to have a relevant joint-spectrum problem.
As such, for the study of the topology in physical system within the spectral localizer framework, no further approximation is required beyond the discretization of the master equation, yielding a potentially more accurate description of a system's topology.
%


\subsection{The photonic master equation}

The master equations for photonic systems are given by Maxwell's equations. Assuming that the materials used are linear and time-independent, the electromagnetic fields can be written in a time-harmonic form $e^{-i \omega t}$ such that a photonic system can be described by the source-free Maxwell's equations
\begin{align}
\label{eq:maxwell_eq_curl_E}
\nabla \times \vec{E}(\vec{x}) 
& = i \omega \bar{\mu}(\vec{x}) \vec{H}(\vec{x})
, \\[1.ex]
\label{eq:maxwell_eq_curl_H}
\nabla \times \vec{H}(\vec{x}) 
& = - i \omega \bar{\epsilon}(\vec{x}) \vec{E}(\vec{x})
, \\[1.ex]
\label{eq:maxwell_eq_div_D}
\nabla \cdot [\bar{\epsilon}(\vec{x}) \vec{E}(\vec{x})]
& = 0
, \\[1.ex]
\label{eq:maxwell_eq_div_B}
\nabla \cdot [\bar{\mu}(\vec{x}) \vec{H}(\vec{x})]
& = 0
,
\end{align}
where 
$\omega$ is the angular frequency,
$\vec{E}(\vec{x})$ and $\vec{H}(\vec{x})$ are the electric and magnetic fields, and
$\bar{\epsilon}(\vec{x})$ and $\bar{\mu}(\vec{x})$ are the permittivity and permeability tensors.
Previously, the spectral localizer framework has been applied to describe the topology of 2D photonic crystals~\cite{Cerjan2022a, Dixon2023, Cerjan2023b}.
These prior studies took advantage of finite-difference discretizations of Maxwell equations [Eqs.~\eqref{eq:maxwell_eq_curl_E}-\eqref{eq:maxwell_eq_curl_H}] to automatically satisfy the divergence-free condition [Eqs.~\eqref{eq:maxwell_eq_div_D}-\eqref{eq:maxwell_eq_div_B}].
However, this approach is not easily scalable, as it requires a uniform mesh and therefore requires very large matrices for the simulation of realistic designs of 3D systems.

Instead, here, we use the finite-element method (FEM) for a more general approach to any physical system described by a master equation.
The key advantage of FEM-localizer framework is that it allows for a much coarser meshing and therefore reduces the size of the matrices involved. Moreover, the method easily incorporates physical processes with different characteristic length scales, where additional equations of motion can be included to described the relevant processes and their couplings.
For example, dispersion can be described by introducing auxiliary equations for the material's internal degrees of freedom responsible for this effect~\cite{Raman2010} that are then coupled to Maxwell's equations [Eqs.~\eqref{eq:maxwell_eq_curl_E}-\eqref{eq:maxwell_eq_div_B}].
To best make use of the FEM approach for photonic systems, here we use the Helmholtz equation for the electric field $\vec{E}(\vec{x})$, derived from Eqs.~\eqref{eq:maxwell_eq_curl_E}-\eqref{eq:maxwell_eq_div_B}
\begin{equation}
\label{eq:master_eq}
\nabla \times \left( \bar{\mu}^{-1}(\vec{x}) \nabla \times \vec{E}(\vec{x}) \right) - \omega^2 \bar{\epsilon}(\vec{x}) \vec{E}(\vec{x}) = 0
,
\end{equation}
as the master equation to probe the topology in photonic systems.
%


\subsection{Overview of the finite-element method}

\begin{figure}[!]
\center
\includegraphics[width=\columnwidth]{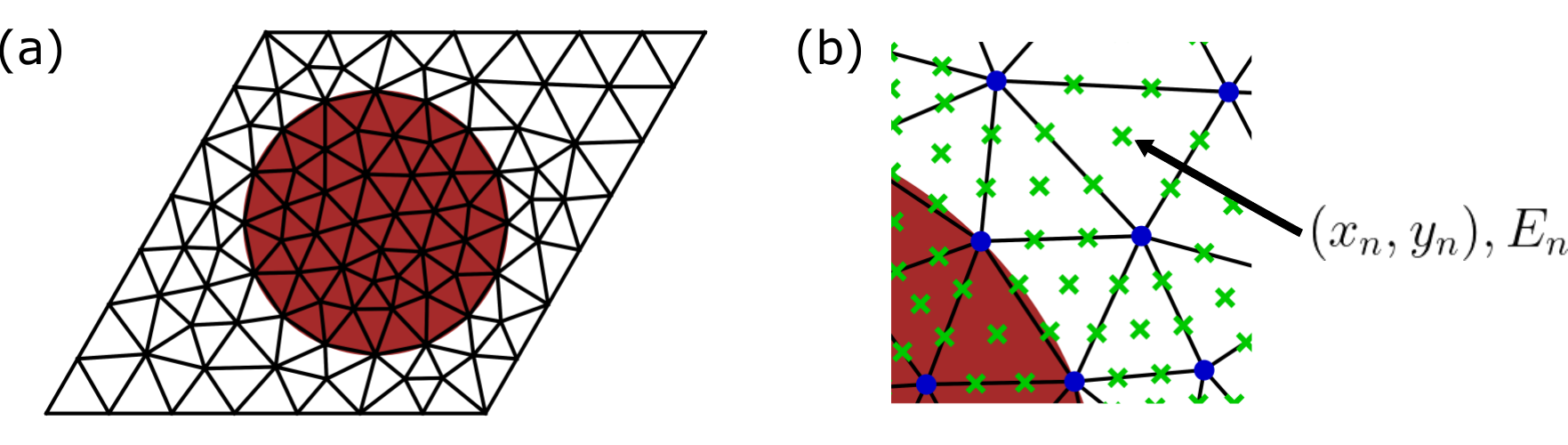}
\caption{
\textbf{Position coordinate of the degrees of freedom used for constructing the position matrices within the finite-element method discretization.}
(a) Schematic of the discretization into finite elements of the two-dimensional (2D) simulation domain composed of a single dielectric rod (red shaded region) at the center of a triangular unit cell.
(b) Zoom-in of (a) where the mesh nodes and extended mesh nodes are depicted using blue and green markers, respectively. 
In this context, the 2D structure is solved for transverse magnetic polarization ($H_z \neq 0$) and the shape function used are the curl elements.
The position of the extended mesh nodes corresponding to the unknown weighting coefficient $E_n$ are located at $(x_n, y_n)$. 
}
\label{fig:fem_dof}
\end{figure}

In this section, we briefly outline FEM discretizations, with a focus on the relevant information for then incorporating the resultant equations of motion into the spectral localizer framework. A more detailed description of FEM can be found in FEM textbooks~\cite{Rylander2013}.

The FEM discretization starts by reformulating the master equation into its weak form, where the differential equation is no longer satisfied exactly at every point of the mesh.
The weak form of the master equation consists in turning the differential equation into an integral equation, via the method of integration by parts, to improve numerical stability.
Indeed, the differentiation of the solutions of the master equation may be limited at the boundaries of the simulation domain or at some material interfaces, where a jump in their values can be observed.
The weak form is therefore obtained by multiplying the master equation with a weight function $\vec{\alpha}(\vec{x})$ and by evaluating the overlap integral over the simulation region $\Omega$.
The weak form for the master equation Eq.~\eqref{eq:master_eq} is then
\begin{multline}
\label{eq:weak_form}
\int_\Omega \left( \bar{\mu}^{-1}(\vec{x}) \nabla \times \vec{E}(\vec{x}) \right) \cdot \left( \nabla \times \vec{\alpha}(\vec{x}) \right) d\vec{x} \\
- \omega^2 \int_\Omega \bar{\epsilon}(\vec{x}) \vec{E}(\vec{x}) \cdot \vec{\alpha}(\vec{x}) d\vec{x} = 0
.
\end{multline}
%
%

The solution vector, $\vec{E}(\vec{x})$ for Eq.~\eqref{eq:master_eq}, is decomposed in terms of the shape functions $\vec{w}_n$ 
\begin{equation}
\vec{E}(\vec{x}) = \sum_n E_n \vec{w}_n(\vec{x})
,
\end{equation}
with unknown weights $E_n$ located along the extended mesh nodes [see for example the green crosses in Fig.~\ref{fig:fem_dof}].
Importantly, these shape functions are chosen to satisfy the desired properties of the system's equation. 
For instance, when considering Maxwell's equations, it is imperative to fulfill both the divergence-free conditions and interface conditions.
In this context, the so-called curl elements~\cite{Nedelec1980} can be chosen as shape functions.
These curl elements inherently satisfy the divergence condition [Eqs.~\eqref{eq:maxwell_eq_div_B}-\eqref{eq:maxwell_eq_div_D}] as they are divergence-free functions.
Furthermore, the curl elements enforce the criterion that the tangential component of the electric field must be continuous while the normal component can be discontinuous across interfaces, in accordance with the interface conditions set by the Maxwell's equations.

Altogether, the solution to the master equation is obtained by solving the FEM-discretized master equation after performing numerical integration, which can be written as a system of linear equations in the matrix-form
\begin{equation}
\label{eq:fem_weak_form_eliminated}
H_\text{eff}(\omega) \vec{\Psi} = \vec{0}
,
\end{equation}
where $\vec{\Psi} = (\ldots, E_n, \ldots)^\text{T}$ is the solution vector, the electric field $\vec{E}(\vec{x})$, at frequency $\omega$.
The magnetic field $\vec{H}(\vec{x})$ can then be derived using Eq.~\eqref{eq:maxwell_eq_curl_E}.
Notably, $H_\text{eff}(\omega)$ is a good candidate to be an effective Hamiltonian for insertion into the spectral localizer as the eigenvector with zero eigenvalue is a solution to the master equation.


\subsection{Incorporating boundary conditions}

Although there are many available implementations of FEM, here we describe the methodology based on the FEM discretization from the commercial software \textsc{COMSOL Multiphysics}~\cite{COMSOL} as this is a widely used software across many different physical platforms. Using details from this specific FEM implementation, we then show how to develop a Hamiltonian that incorporates the system's boundary conditions.

Using the \textbf{Eigenvalue Solver} algorithm in \textsc{COMSOL}, the FEM discretization leads to solving the following set of equations for the solution vector in the extended mesh $\vec{\Psi}$, 
\begin{align}
\label{eq:eigenvalue_solver_comsol_H}
H_\text{eff}(\omega) \vec{\Psi} + N_F \vec{\Lambda} = \vec{0}
, \\[1.ex]
\label{eq:eigenvalue_solver_comsol_N}
N \vec{\Psi} = \vec{0}
,
\end{align}
where $N_F$ and $N$ are respectively the constraint force Jacobian matrix and the constraint Jacobian matrix, and $\vec{\Lambda}$ is a vector made of the Lagrange multipliers for the boundary conditions and fictitious degrees of freedom. 
In Eqs.~\eqref{eq:eigenvalue_solver_comsol_H}-\eqref{eq:eigenvalue_solver_comsol_N}, $H_\text{eff}$ is a matrix-valued function
\begin{equation}
H_\text{eff} : \omega \rightarrow (-i \omega)^2 M - (-i \omega) C + K
,
\end{equation}
where $\omega \in \mathbb{C}$,
$M$ is the mass matrix, $C$ is the damping matrix, and $K$ is the stiffness matrix.
The real part of $\omega$ corresponds to the solution's angular frequency while its imaginary part is its decay rate.
In sum, Eqs.~\eqref{eq:eigenvalue_solver_comsol_H}-\eqref{eq:eigenvalue_solver_comsol_N} contain the discretized weak form of the master equation [see the term $H_\text{eff}(\omega) \vec{\Psi}$ in Eq.~\eqref{eq:eigenvalue_solver_comsol_H}], as well as additional terms that account for the boundary conditions.

In order to remove the additional terms that incorporate the boundary conditions and solve an equation that resembles Eq.~\eqref{eq:fem_weak_form_eliminated}, the constraint equation is first solved.
Namely, Eq.~\eqref{eq:eigenvalue_solver_comsol_N} is solved as
\begin{equation}
N \vec{\Psi}_d = \vec{0}
\end{equation}
with $\vec{\Psi}_d = \vec{0}$.
Then
\begin{equation}
\label{eq:solution_Psi}
\vec{\Psi} = \left( \vec{\Psi}_d + \text{Null} \vec{\Psi}_c \right)
\end{equation}
is a solution of Eqs.~\eqref{eq:eigenvalue_solver_comsol_H}-\eqref{eq:eigenvalue_solver_comsol_N}.
Equations~\eqref{eq:eigenvalue_solver_comsol_H}-\eqref{eq:eigenvalue_solver_comsol_N} are therefore reduced to find $\vec{\Psi}_c$, the solution of the eliminated matrix equation
\begin{equation}
\label{eq:eliminated_H}
H_{\text{eff},c}(\omega) \vec{\Psi}_c = \vec{0}
\end{equation}
where $H_{\text{eff},c}$ is the matrix-valued function defined as
\begin{equation}
H_{\text{eff},c}(\omega) = \text{Nullf}^T H_\text{eff}(\omega) \text{Null}
\end{equation}
with
\text{Null} and \text{Nullf} being composed of basis vectors spanning the null space of $N$ and $N_F^\text{T}$, respectively,
\begin{align}
N \text{Null} = \vec{0}
, \\[1.ex]
\text{Nullf}^T N_F = \vec{0}
.
\end{align}
Physically, the eliminated matrix equation [Eq.~\eqref{eq:eliminated_H}] considers the eliminated effective Hamiltonian $H_{\text{eff},c}(\omega)$ where all the degrees of freedom involved in the boundary conditions have been accounted for and removed.
Consequently, $H_{\text{eff},c}(\omega)$ is the effective Hamiltonian compatible with the spectral localizer framework, and not $H_\text{eff}(\omega)$.
%


\subsection{The spectral localizer via finite-element method}

As the eigenvector of $H_{\text{eff},c}(\omega)$ with eigenvalue zero is physically meaningful because it corresponds to a solution of eliminated Maxwell's equations, $H_{\text{eff},c}(\omega)$ can be chosen to be the effective Hamiltonian matrix in the spectral localizer framework.
Accordingly, the position matrices $X_{j,c}$ need to be constructed in the same vector space as $H_{\text{eff},c}(\omega)$.
In the non-eliminated vector space, the position matrices $X_j$ are diagonal matrices with entries corresponding to the $j$-th coordinate position of the $n$-th FEM degrees of freedom $x_{j,n}$, i.e.\ the positions of the extended mesh nodes [see for example the green crosses in Fig.~\ref{fig:fem_dof}],
\begin{equation}
X_j = 
\left(
\begin{array}{ccc}
\ddots &  &  \\ 
 & x_{j,n} &  \\ 
 &  & \ddots
\end{array} 
\right)
.
\end{equation}
The position matrices in the eliminated space $X_{j,c}$ are therefore obtained by projecting $X_j$ onto the eliminated space via $\text{Null}$ and $\text{Nullf}$ as
\begin{equation}
X_{j,c} = \text{Nullf}^T X_j \text{Null}
.
\end{equation}

Given the unique mathematical and computational complexities associated with using a FEM, there are several modifications that must be made to the spectral localizer framework [Eq.~\eqref{eq:localizer}] to probe the systems local topology.
First, to probe the topology at spatial position coordinate $(\ldots, x_j, \ldots)$, $x_j$ needs to be expressed in the eliminated space in accordance to the position matrix $X_{j,c}$. 
This can be implemented by directly projecting $(X_j - x_j I)$, appearing in Eq.~\eqref{eq:localizer}, onto the eliminated space, $(X_j - x_j I)_c = \text{Nullf}^T (X_j - x_j I) \text{Null}$, and by probing at spatial position zero namely $\vec{\lambda}=(0, \ldots, 0, E)$.
Second, as only the zero eigenvalue of $H_{\text{eff},c}$ corresponds to a physically meaningful solution of the Maxwell's equations, the localizer has to be probed at $\vec{\lambda}=(\ldots, 0)$, namely the effective Hamiltonian $H_\text{eff}(\omega)$ itself carries the information about the frequency $\omega$ at which the system's topology is being classified.
Thus, for shorthand notation, the topology at location $(x_1, \dots, x_d, \omega)$ is probed using the spectral FEM-localizer defined as 
\begin{align}
\begin{split}
& \hat{L}_{(x_1, \ldots, x_d, \omega)}(X_{1,c}, \ldots, X_{d,c}, H_{\text{eff},c}) = \\ 
& L_{(0, \ldots, 0, 0)}(X_1 - x_1 I_c, \ldots, X_d - x_d I_c, H_{\text{eff},c}(\omega)) 
,
\end{split}
\end{align}
with
\begin{equation}
I_c = \text{Nullf}^T \text{Null}
.
\end{equation}
For example, for a 2D Hermitian system, the spectral FEM-localizer can be written using the Pauli spin matrices as the choice of Clifford representation, yielding
\begin{align}
\label{eq:fem_localizer_2d}
\begin{split}
& \hat{L}_{(x, y, \omega)}(X_c, Y_c, H_{\text{eff},c}) = \\
& 
\left(
\begin{array}{cc}
H_{\text{eff},c}(\omega) & \kappa (X_c - x I_c) - i\kappa(Y_c - y I_c) \\ 
\kappa (X_c - x I_c) + i\kappa(Y_c - y I_c) & -H_{\text{eff},c}(\omega)
\end{array}
\right)
,
\end{split}
\end{align}
and the local topological marker, for class A systems, at position $(x,y)$ and angular frequency $\omega$ is obtained from
\begin{multline}
\label{eq:fem_local_chern_nb}
C_{(x,y,\omega)}^{\textrm{L}}(X_c, Y_c, H_{\text{eff},c}) = \\
\frac{1}{2} \textrm{sig} \big[ \hat{L}_{(x, y, \omega)}(X_c, Y_c, H_{\text{eff},c}) \big]
.
\end{multline}

Altogether, the effective Hamiltonian $H_{\text{eff},c}(\omega)$ solves the system's master equation rigorously with the only approximation being the discretization, taking into account all the possible processes in the physical system.
The retained information in the effective Hamiltonian therefore gives us a more rigorous description of the topology in the physical system.
For instance, in photonic systems, this approach can directly incorporate the radiative processes overlooked in the literature by using the non-Hermitian line-gap extension of the spectral localizer~\cite{Cerjan2023}.
Instead of using the Hermitian localizer [Eq.~\ref{eq:fem_localizer_2d}], the non-Hermitian spectral localizer for classifying 2D non-Hermitian (lossy) systems is now written as
\begin{align}
\label{eq:fem_localizer_2d_nh}
\begin{split}
& \hat{L}^{(\text{NH})}_{(x, y, \omega)}(X_c, Y_c, H_{\text{eff},c}) = \\
& 
\left(
\begin{array}{cc}
H_{\text{eff},c}(\omega) & \kappa (X_c - x I_c) - i\kappa(Y_c - y I_c) \\ 
\kappa (X_c - x I_c) + i\kappa(Y_c - y I_c) & -H_{\text{eff},c}(\omega)^\dagger
\end{array}
\right)
,
\end{split}
\end{align}
and the local topological marker at position $(x,y)$ and angular frequency $\omega$ is obtained from
\begin{multline}
\label{eq:fem_local_chern_nb_nh}
C_{(x,y,\omega)}^{\textrm{L}, (\text{NH})}(X_c, Y_c, H_{\text{eff},c}) = \\
\frac{1}{2} \textrm{sig}_\mathbb{R} \big[ \hat{L}^{(\text{NH})}_{(x, y, \omega)}(X_c, Y_c, H_{\text{eff},c}) \big]
.
\end{multline}
where $\text{sig}_\mathbb{R}$ denotes the matrix's difference between its number of positive and negative eigenvalues with respect to their real part.
Additionally, the localizer gap becomes
\begin{multline}
\label{eq:localizer_gap_nh}
\mu^{\text{C}, (\text{NH})}_{(x,y\omega)}(X_c, Y_c, H_{\text{eff},c}) = \\ 
\min \big[ | \text{Re} \{ \sigma( \hat{L}^{(\text{NH})}_{(x, y, \omega)}(X_c, Y_c, H_{\text{eff},c}) ) \} | \big]
.
\end{multline}
Notably, $\omega$ can be complex in the lossy system and can include the damping term for calculating the local markers.
However, by using the line-gap extension of the spectral localizer for 2D class A systems, the imaginary part of $\omega$ should not matter in the calculation of the topology~\cite{Cerjan2023}.

Finally, it is emphasized here that the matrices for constructing the spectral localizer can be readily obtained from \textsc{COMSOL} once the \textbf{Eigenvalue Solver} study has been run, regardless of the module used: the matrices $M, C, K$, $\text{Null}$, and $\text{Nullf}$ for determining $H_{\text{eff},c}$ and the spatial position $(\ldots, x_j, \ldots)$ of the extended mesh nodes can be directly accessed from the \textsc{COMSOL} functions in \textsc{MATLAB}~\cite{MATLAB}.
The spectral FEM-localizer framework can therefore be immediately applied to the wealth of examples and designs considered by the \textsc{COMSOL} community.


\section{Example of 2D photonic Chern structures}

\begin{figure}[t!]
\center
\includegraphics[width=\columnwidth]{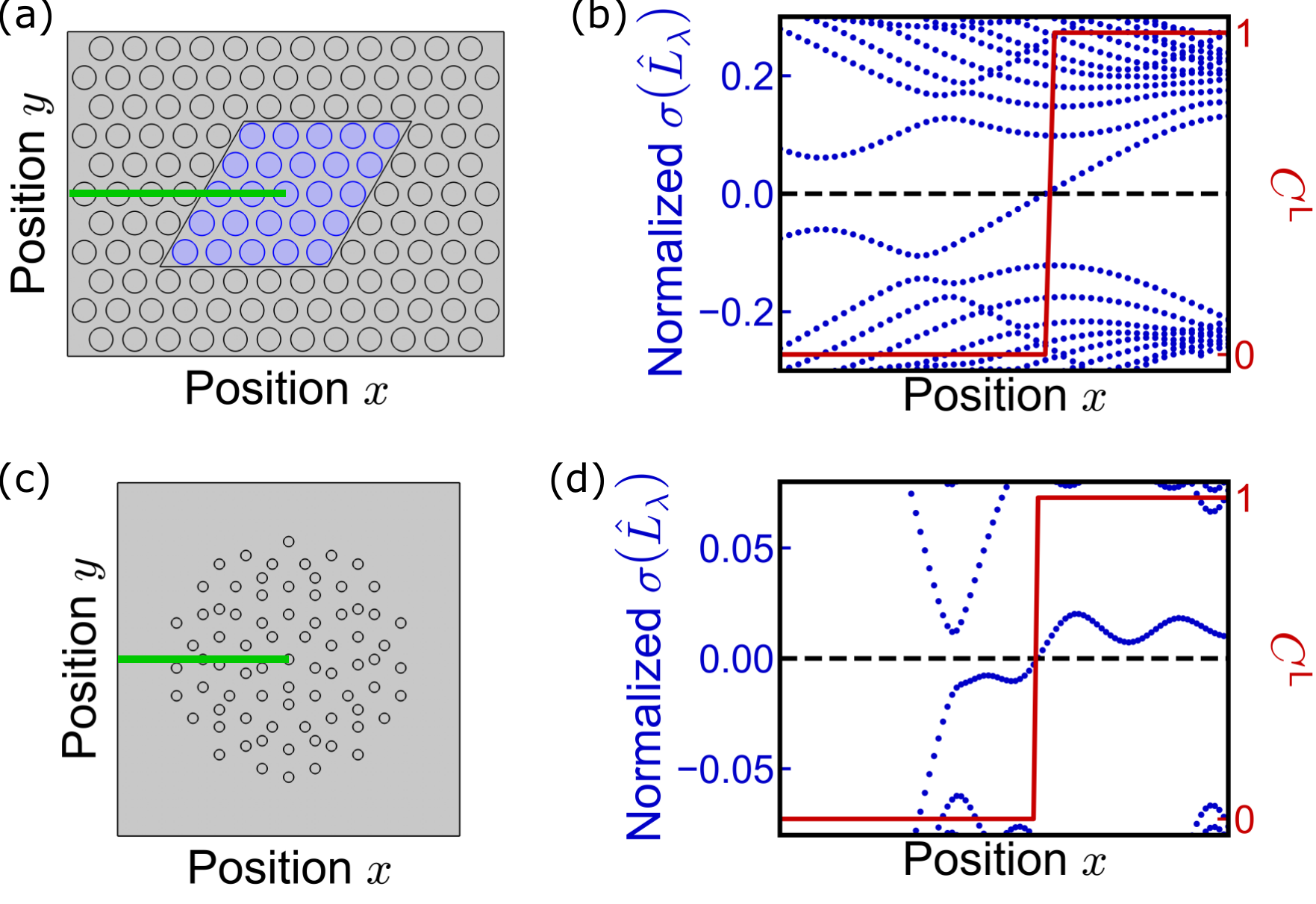}
\caption{
\textbf{Probing of the local topology in two-dimensional photonic Chern system examples.}
(a) Design of the 2D photonic Chern heterostructure. 
The inner parallelogram is a non-trivial topological lattice with lattice constant $a=\SI{1}{\micro\metre}$ made of dielectric rods, $\bar{\epsilon}_{jj}=14$ for $j=x,y,z$, with radius $r=0.37a$ embedded in a gyro-electric background, $\bar{\epsilon}_{jj} = 1$, $\bar{\epsilon}_{xy} = -0.4i$, that breaks time-reversal symmetry.
The outer lattice is a topologically trivial lattice with lattice constant $a=\SI{1}{\micro\metre}$ composed of air rods, $\bar{\epsilon}_{jj}=1$, with radius $r=0.35a$ in a dielectric background with $\bar{\epsilon}_{jj}=5.5$.
(b) Spectrum of the FEM-localizer $\sigma(\hat{L}_{(x,y_0,\omega_0)})$ normalized by $10^{-4} \norm{H_{\text{eff},c}(\omega_0)}$ and the local Chern number $C_{(x,y_0,\omega_0)}^{\textrm{L}}$ along the green line in (a) at $y_0 = 0$ and frequency $\omega_0 = 0.37 [2 \pi c/a]$, with $\kappa = 1.5 \big[ 10^{-4} \norm{H_{\text{eff},c}(\omega_0)}/\norm{X} \big]$.
(c) Design of the photonic Chern quasicrystal based on a Penrose tiling.
The dielectric rods are located at the vertices of the Penrose tiling in a gyro-electric background, $\bar{\epsilon}_{jj} = 1$, $\bar{\epsilon}_{xy} = -0.4i$.
(d) Spectrum of the FEM-localizer $\sigma(\hat{L}_{(x,y_0,\omega_0)})$ normalized by $10^{-4} \norm{H_{\text{eff},c}(\omega_0)}$ and the local Chern number $C_{(x,y_0,\omega_0)}^{\textrm{L}}$ along the green line in (a) at $y_0 = 0$ and frequency $\omega_0 = 0.37 [2 \pi c/a]$, with $\kappa = \big[ 10^{-4} \norm{H_{\text{eff},c}(\omega_0)}/\norm{X} \big]$.
}
\label{fig:example_chern}
\end{figure}

As an initial demonstration of the versatility of the spectral FEM-localizer framework, we consider two fundamental examples in topological photonics~[Fig.~\ref{fig:example_chern}]:
a 2D Haldane photonic crystal heterostructure~\cite{Haldane2008} that is the canonical photonic Chern insulator, and a 2D photonic quasicrystal~\cite{Zhang2023} that is an aperiodic system where topological band theory cannot be applied.
In both cases, the photonic system is first described by the full-wave Helmholtz equation (using the \textit{Wave Optics} module) and then discretized using FEM via \textsc{COMSOL Multiphysics}.
Finally, the spectral FEM-localizer is constructed from the \textbf{Eigenvalue Solver} study and used to classify the systems' topology.

\subsection{2D Photonic Chern crystal}

For a first example, we focus on the 2D Haldane photonic heterostructure~\cite{Haldane2008} shown in Fig.~\ref{fig:example_chern}(a).
Here, the Haldane heterostructure is made of two triangular lattices of lattice constant $a = \SI{1}{\micro\metre}$ with perfect electric conductor (PEC) boundary conditions, and we are considering the topology of the transverse magnetic modes ($H_z \neq 0$).
The inner lattice is a topologically non-trivial insulator composed of dielectric rods, $\bar{\epsilon}_{jj}=14$ for $j=x,y,z$, with radius $r=0.37a$ embedded in a gyro-electric background, $\bar{\epsilon}_{jj} = 1, \bar{\epsilon}_{xy} = -0.4i$, to break time-reversal symmetry.
The outer lattice is a topological trivial insulator composed of air rods, $\bar{\epsilon}_{jj}=1$, of radius $r=0.35a$ in a dielectric background with $\bar{\epsilon}_{jj}=5.5$.

The spectral FEM-localizer and local Chern marker in Eq.~\eqref{eq:fem_localizer_2d} and Eq.~\eqref{eq:fem_local_chern_nb}, respectively, can be used to diagnose the topology of this lossless 2D system.
However, one should note that the eliminated effective Hamiltonian $H_{\text{eff},c}$ is non-Hermitian due to the projection onto the eliminated space.
Nevertheless, the non-Hermitian part is found to be negligible and only the Hermitian part is kept in $H_{\text{eff},c}$ as Hermitian materials and lossless boundary conditions have been used.
The spectrum of the spectral FEM-localizer $\sigma(\hat{L}_{(x,y_0,\omega_0)})$ and the local Chern number $C_{(x,y_0,\omega_0)}^{\textrm{L}}$ along the path depicted by the green line in Fig.~\ref{fig:example_chern}(a) at $y_0 = 0$ and frequency $\omega_0 = 0.37 [2 \pi c/a]$ are shown in Fig.~\ref{fig:example_chern}(b), demonstrating the local topological picture of the heterostructure.
As expected from topological band theory, inside the topological band gap at around $\omega_0 = 0.37 [2 \pi c/a]$, the inner lattice is topological with $C^\text{L}=1$ while the outer lattice is trivial with $C^\text{L}=0$.
Therefore, the spectral FEM-localizer correctly captures the change of topology as demonstrated by the eigenvalue crossing with respect to zero of the spectrum of $L$ near the heterostructure's interface.
%


\subsection{2D Photonic Chern quasicrystal}

As a second example, we investigate the topology of a magnetooptic 2D photonic quasicrystal surrounded by a homogeneous material, again for transverse magnetic modes ($H_z \neq 0$).
Notably, this example cannot be classified using topological band theory as the system is not periodic and therefore does not possess a band structure.
Moreover, the quasicrystal is surrounded by a homogeneous material that is gapless rather than gapped (or insulating), yielding an ill-defined notion of bulk topological invariant and topological robustness.
The topological quasicrystal is constructed from a Penrose tiling~\cite{DeBruijn1981} where dielectric rods are positioned on the vertices of the tiling~\cite{Zhang2023}, as shown in Fig.~\ref{fig:example_chern}(c).
The photonic quasicrystal is composed of dielectric rods with permittivity $\bar{\epsilon}_{jj} = 14$ and radius $r=\SI{0.13}{\micro\metre}$, embedded in a gyro-electric background, $\bar{\epsilon}_{jj} = 1, \bar{\epsilon}_{xy} = -0.4i$ to break time-reversal symmetry, and PEC boundary conditions are used.

Similar to the crystalline example, the spectral FEM-localizer and local marker in Eq.~\eqref{eq:fem_localizer_2d} and Eq.~\eqref{eq:fem_local_chern_nb} are used to probe the topology of this photonic system once the non-Hermitian part is removed from the eliminated effective Hamiltonian $H_{\text{eff},c}$.
Figure~\ref{fig:example_chern}(d) shows the spectrum of the FEM-localizer and the local Chern number as the probe location is varied, revealing the topology despite the system's aperiodicity and the lack of a surrounding insulator.
In particular, there is a crossing of the spectrum of the localizer near the boundary of the structural interface, and thus a change of the local Chern number as the probe coordinate is moved from the trivial homogeneous material to the center of the quasicrystal along the green line depicted in Fig.~\ref{fig:example_chern}(c), and at $\omega_0 = 0.37 [2 \pi c/a]$.
A similar plot for the spectrum of the spectral localizer and the local Chern number can be realized along the frequency axis, as shown in the Supplemental Material~\cite{supp}, demonstrating some range of frequencies $\omega$ for which the quasicrystal is topologically non-trivial.  
As such, this example highlights how the spectral FEM-localizer is capable to identify the topology of the system without the need of a bandstructure or a bulk band gap.
%


\section{Photonic Chern slab}

\begin{figure}[t!]
\center
\includegraphics[width=\columnwidth]{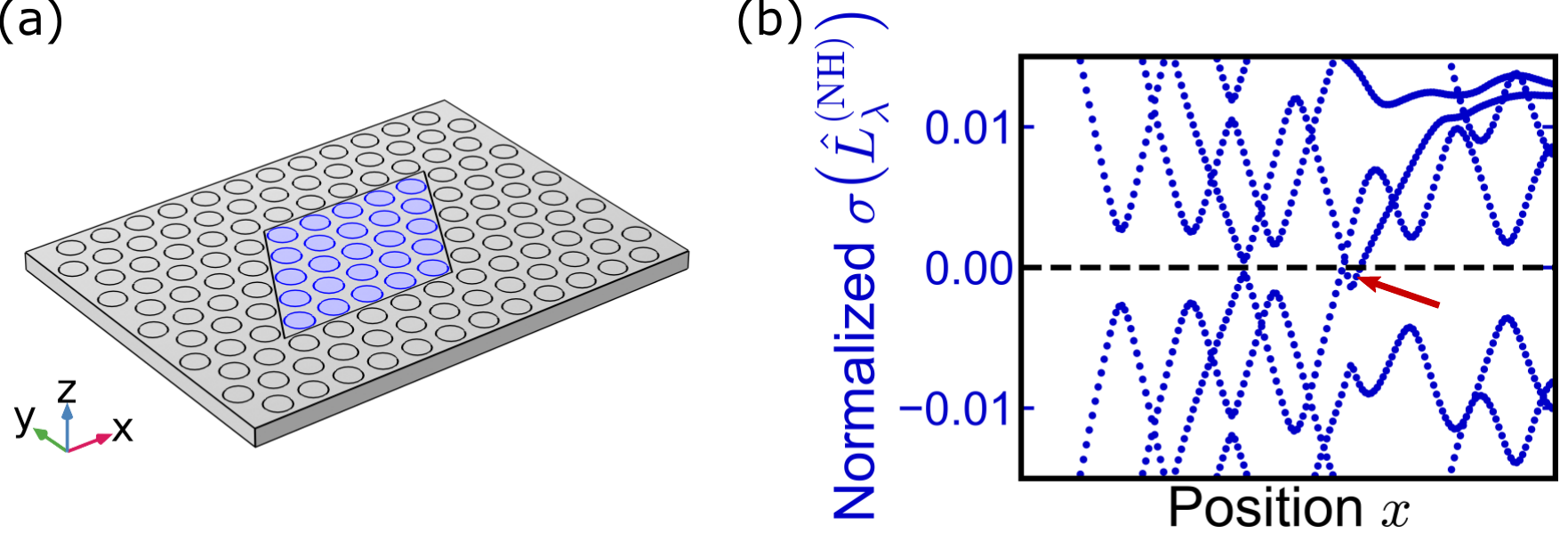}
\caption{
\textbf{Probing of the local topology in a photonic Chern slab system.}
(a) Design of the photonic Chern slab. 
The inner parallelogram is a non-trivial topological lattice with lattice constant $a=\SI{1}{\micro\metre}$ made of dielectric rods, $\bar{\epsilon}_{jj}=14$ for $j=x,y,z$, with radius $r=0.37a$ and height $t = 0.5a$ embedded in a gyro-electric slab, $\bar{\epsilon}_{jj} = 1$, $\bar{\epsilon}_{xy} = -0.4i$, with thickness $t = 0.5a$.
The outer lattice is a topological trivial lattice with lattice constant $a=\SI{1}{\micro\metre}$ composed of air rods, $\bar{\epsilon}_{jj}=1$, with radius $r=0.35a$ and height $t = 0.5a$ in a dielectric slab, $\bar{\epsilon}_{jj}=5.5$, with thickness $t$.
(b) Spectrum of the FEM-localizer $\sigma(\hat{L}^{(NH)}_{(x,y_0,\omega_0)})$ normalized by $10^{-4} \norm{H_{\text{eff},c}(\omega_0)}$ along the green line in (a) at $y_0 = 0$ and frequency $\omega_0 = 0.37 [2 \pi c/a]$, with $\kappa = 1.5 \big[ 10^{-4} \norm{H_{\text{eff},c}(\omega_0)}/\norm{X} \big]$.
}
\label{fig:chern_slab}
\end{figure}

Photonic structures are inherently 3D, and as such the 2D photonic designs extensively studied in the literature and described in the previous section are only approximations to actual photonic slabs realized experimentally. While there are standard methods for developing 3D photonic crystal slabs whose resonance band structures are quantitatively similar to the band structures of 2D photonic crystals~\cite{Joannopoulos2011}, these methods only approximate the Hermitian portion of the band structure, not the radiative portion.
Moreover, approximating radiative losses due to a system's environment as material absorption within the system is uncontrolled, as this is fundamentally relocating degrees of freedom that were outside of a structure to be inside of it. Indeed, this is a particularly problematic approximation for topological systems, whose primary features are boundary-localized states---relocating degrees of freedom changes what it means for a state or resonance to be localized.
Therefore, out-of-plane radiative losses, an inherent aspect of photonic slabs, have been neglected in previous theoretical treatments of topological photonics, as the radiative loss cannot be accounted for using a band theoretic approach.
Nevertheless, as we have already seen in the case of the quasicrystal, the spectral FEM-localizer allows to diagnose the topology beyond the scope of topological band theory.
Here, we show that the spectral FEM-localizer can be used to classify the topology of photonic crystal slabs and directly incorporate out-of-plane radiative losses using radiative boundary conditions.

As an example illustrating the probing of the topology in photonic slab while accounting for the out-of-plane radiative loss, we consider a free-standing photonic crystal slab embedded in air, as shown in Fig.~\ref{fig:chern_slab}(a). 
This is the 3D slab version of the Haldane photonic heterostructure studied in Fig.~\ref{fig:example_chern}(a), with PEC on the $x$- and $y$-boundaries and a radiative boundary condition implemented through perfectly matched layers (PML) on the $z$-boundaries.
The parameters used are the same as in the Chern heterostructure example [Fig.~\ref{fig:example_chern}(a)] except that now the background is a slab of thickness $t=0.5a$ embedded in air, and the rods have the same finite height $t = 0.5a$. 

As the system is 3D and in class A in the Altland-Zirnbauer classification~\cite{Altland1997, Schnyder2008, Kitaev2009, Ryu2010}, the topology for the 2D topological edge state in the slab can be classified using an integer invariant such as the Chern number.
Within the spectral localizer framework, the topology is diagnosed by squeezing the $z$-direction into a single value (i.e., all mesh vertices are retained, but their coordinates is reduced $(x,y,z) \rightarrow (x,y)$), performing a version of dimensional reduction to enable the calculation of a strong 2D invariant of a 3D system.
Physically, this is equivalent as looking at the change of topology as we move in the $(x,y)$-plane, irrespective of the z-coordinate.
Despite the squeezing of the $z$-direction, all the information from the 3D geometry, including the out-of-plane radiative loss, is retained for the assessment of the topology as the effective Hamiltonian $H_{\text{eff},c}$ is derived from the full 3D geometry.
Thus, the 2D non-Hermitian FEM-localizer in Eq.~\ref{eq:fem_localizer_2d_nh} can be used to identify the topology in the slab with radiative boundary conditions, giving an accurate probing of the topology in the photonic slab where all the possible processes are accounted for.

The topology is studied by looking at the spectrum of the spectral FEM-localizer $\sigma(\hat{L}^{(NH)}_{(x,y_0,\omega_0)})$ along the green path in Fig.~\ref{fig:chern_slab}(a) at $y_0 = 0$, and at the (incomplete) band gap around $\omega_0 = 0.37 [2 \pi c/a]$, as shown in Fig.~\ref{fig:chern_slab}(b).
The plot demonstrates a net crossing in the spectrum with respect to zero [see red arrow in Fig.~\ref{fig:chern_slab}(b)], indicating a change of topology near the boundary between the outer and inner lattices, similar to what is observed in Figs.~\ref{fig:example_chern}(a)-(b).
However, the topological protection given by the localizer gap now takes into account the radiative losses of the topological edge slab state, and as such the protection is weaker than would be predicted from the 2D band structure.
%


\section{Conclusion}

In conclusion, using the operator-based approach of the spectral localizer, we have developed a general framework for studying the topology in realistic photonic structures directly from the discretized master equations of the system using finite-element methods (FEM).
In particular, we studied the topology in photonic systems derived directly from the full-wave Maxwell equations.
Using the photonic Chern insulators and the photonic Chern quasicrystal, we have demonstrated the ability of the proposed spectral FEM-localizer framework to correctly capture the local topology in photonic topological materials.
Moreover, the framework have been applied to a photonic Chern slab predicting genuine topological protection of the topological edge slab state when taking into account possible radiative loss of the slab state. 
Looking forward, we expect that the spectral FEM-localizer's ability to classify the topology of photonic systems will be useful for developing next-generation devices, and we anticipate the generality of the framework to be of practical use for tackling topological problems in other complex physical platforms such as in acoustic systems~\cite{Ma2019}, plasmonic systems~\cite{Raman2010, Pocock2018, Proctor2020}, and in polaritonic systems~\cite{Karzig2015, Klembt2018, Septembre2023}.




\section*{Acknowledgments}
We acknowledge discussions with L. Zhang and O. Miller.
S.W.\ acknowledges support from the Laboratory Directed Research and Development program at Sandia National Laboratories. 
T.A.L.\ acknowledges support from the National Science Foundation, Grant No. DMS-2110398. 
A.C.\ acknowledges support from the U.S.\ Department of Energy, Office of Basic Energy Sciences, Division of Materials Sciences and Engineering.
This work was performed in part at the Center for Integrated Nanotechnologies, an Office of Science User Facility operated for the U.S. Department of Energy (DOE) Office of Science.
Sandia National Laboratories is a multimission laboratory managed and operated by National Technology \& Engineering Solutions of Sandia, LLC, a wholly owned subsidiary of Honeywell International, Inc., for the U.S. DOE's National Nuclear Security Administration under Contract No. DE-NA-0003525. 
The views expressed in the article do not necessarily represent the views of the U.S. DOE or the United States Government.

\bibliography{main_v0}

\end{document}